\documentclass{PoS}
\title{U(1) Vacuum, Chern-Simons Diffusion and Real-time Simulations}

\usepackage{tikz}
\usepackage{pgfplots}
\usetikzlibrary{positioning}
\usepgfplotslibrary{fillbetween}
\usepgfplotslibrary{patchplots}
\usetikzlibrary{arrows,shapes,positioning}
\usetikzlibrary{pgfplots.groupplots}
\usepackage{amsmath, amssymb, mathrsfs}
\newcommand{\mn}{\ensuremath{{\mu\nu}}}

\usepgfplotslibrary{external}
\definecolor{shadedsl}{RGB}{180, 180, 180}

\usepgfplotslibrary{external}
\tikzset{external/system call={pdflatex --shell-escape
		\tikzexternalcheckshellescape --halt-on-error --interaction=batchmode
		--jobname "\image" "\texsource"}}
\tikzset{cross/.style={cross out, draw=black, minimum size=2*(#1-\pgflinewidth), inner sep=0pt, outer sep=0pt},
	cross/.default={2.2pt}}
\usetikzlibrary{pgfplots.groupplots}

\ShortTitle{U(1) Vacuum, Chern-Simons Diffusion and Real-time Simulations}

\author{\speaker{Adrien Florio}\\
        Institute of Physics, Laboratory of Particle Physics and Cosmology, Ecole Polytechnique F\'ed\'erale de Lausanne, CH-1015 Lausanne, Switzerland\\
        E-mail: \email{adrien.florio@epfl.ch}}

\abstract{Despite its importance for subjects ranging from cosmology to plasma physics, first principle simulations of the dynamics associated with a $U(1)$ chiral anomaly have been
started only recently. In this work, we report on the current status of these investigations. We discuss a possible set-up and highlight some results. We present a determination of the Chern-Simons
diffusion rate, which shows some discrepancy with the usual effective description which is anomalous magnetohydrodynamic. We also present some exploratory results on the behaviour of an initial chiral chemical potential.}

\FullConference{The 36th Annual International Symposium on Lattice Field Theory - LATTICE2018\\
		22-28 July, 2018\\
		Michigan State University, East Lansing, Michigan, USA.}

\begin{document}

\section{Introduction}

Within the Standard Model, processes mediated through the chiral anomaly produce a rich physics, textbooks examples \cite{9780521670548} being the explanations of the rate $\pi^0\to \gamma\gamma$  or the reason why the $\eta'$ is much more massive than the pions. The dynamics of these processes is also important, with examples ranging from baryogenesis \cite{Kuzmin1985}
to chiral magnetic effect \cite{Kharzeev2016}. While the dynamics of non-abelian anomalous processes has been extensively studied, see \cite{DOnofrio2012a} for an up-to-date computation of the standard model $SU(2)$ sphaleron rate, a first principles study of the anomaly dynamic in the abelian sector is still lacking.  A first step in this direction was taken in \cite{Figueroa2018,Figueroa2018a}. This work,
which mostly reports on \cite{Figueroa2018Boh}, is another one.


The model we investigate is based on scalar electrodynamics coupled to a massless vector-like fermion field $\Psi$
\begin{equation}
\mathcal{L} = -{1\over4} F_{\mu\nu}F^{\mu\nu} - {\bar\Psi}\gamma^\mu D_\mu\Psi - (D_\mu\phi)^*(D^\mu\phi) - V(\phi)
\label{cl}
\end{equation}
with $F_\mn$ the field strength tensor of our gauge field $A_\mu$, $D_\mu=\partial_\mu - i e A_\mu$ and $V(\phi) = m^2|\phi|^2 + \lambda|\phi|^4$. Taking the mass to be positive, we can tune it so that this action becomes a toy-model for the hypercharge sector of the standard model close to the phase transition.
The classical invariance  under chiral rotations is anomalous at the quantum level

\begin{align}
\partial_\mu J_5^\mu &=\frac{e^2}{8\pi^2} F_\mn \tilde{F}^\mn=N_f\partial_\mu K^\mu
\label{eq:an_rel}
\end{align}
with $J^\mu_5={\bar\Psi}\gamma^\mu\gamma_5\Psi$, $\tilde{F}_\mn=\frac{1}{2}\epsilon_{\mu\nu\rho\sigma}F^{\rho\sigma}$ is the hodge-dual of $F_\mn$, $N_f=2$ is the number of flavours and $K^\mu=\frac{e^2}{8\pi^2}\epsilon^{\mu\nu\rho\sigma}A_{\nu}\partial_{\rho}A_{\sigma}$
is the CS-form. In particular, we see that the CS number density  $n_{CS}=K^0$ is nothing else than what is often referred to as the magnetic helicity density
\begin{equation}
n_{CS}=\frac{e^2}{8\pi^2} \vec{A}\cdot\vec{B}
\end{equation}
In particular, for a homogeneous chiral current, the anomaly equation \eqref{eq:an_rel} reduces to
\begin{align}
\partial_0 J_5^0 &=N_f\partial_\mu K^\mu
\end{align}

This model provides us with a rich dynamics which we wish to study with the help of real-time simulations. In section \ref{sec:setup}, we sum-up the set-up which allows us to do so. In section \ref{sec:results}, we discuss results on the Chern-Simons (CS) diffusion rate and present an initial study of the chiral chemical potential dynamics. Section \ref{sec:outlooks} allows us to present some outlooks.

\section{Theoretical Set-Up and Lattice Implementation}
\label{sec:setup}

In this section we sumarise and highlight some of the essential parts of \cite{Figueroa2018,Figueroa2018a}, where our set-up is fully explained.

In the case of a homogeneous chiral current, we can write an effective description without fermions. The anomaly is a non-conservation of chirality and upon integrating out the fermions may be represented as a (homogeneous in our case) chemical potential $\mu_5$ which sources the CS number \cite{Rubakov1986}.
Then, the anomaly becomes an equation for $\mu_5$, which reads
\begin{equation}
\frac{d\mu_5}{dt} = \frac{\sqrt{3}e^2}{8T\pi^2} {1\over V}\int d^3x\,F_\mn \tilde{F}^\mn\,.
\label{mu5eq}
\end{equation}
with $T$ the temperature and $V$ the volume of the system. The equations of motion together with the anomaly  equation can be derived from the following effective action

\begin{equation}
S_{eff}= -\int d^4x\left(- (D_\mu\phi)^*(D^\mu\phi) - V(\phi) + {1\over 4}F_\mn F^\mn - {1\over 2}(\partial_0 a)^2  \right .\\
\left . - {\sqrt{3}e^2\over 8\pi^2}{ a\over T}F_\mn \tilde{F}^\mn\right)\,
\end{equation}
upon identifying $\partial_0 a= {\mu_5\over V}$. Note that this is nothing else that the effective action of a homogeneous axion field  coupled to a $U(1)$ sector.

Having gotten rid of the fermions, putting the theory on the lattice to perform real-time simulations becomes easier. Still, it needs to be done with some care, especially when considering the
CS number \cite{Moore1996a}. One of the perks of an abelian theory over a non-abelian one is that it admits a discretisation directly in terms of gauge potentials and not only in terms of parallel transporters (or links); it admits a non-compact discretisation on top of the usual compact one.
This is the key point which allows us to write down what we will call a simple  lattice topological version of the topological charge density $F^\mn\tilde{F}_\mn$ and of the CS number. We call it topological in the sense that we can construct a discrete  topological charge density which can be written as the discrete divergence
of a discrete CS. Details are presented in \cite{Figueroa2018a}, here we will summarise the basic constituants needed to formulate our problem, which discrete set of equations we are solving and how we are putting everything together to solve our real time problem.

\begin{figure}
  \def\xori{0}
  \def\yori{0}
  \def\xshift{7.3}
  \def\yshift{-0.7}
  \centering

\includegraphics{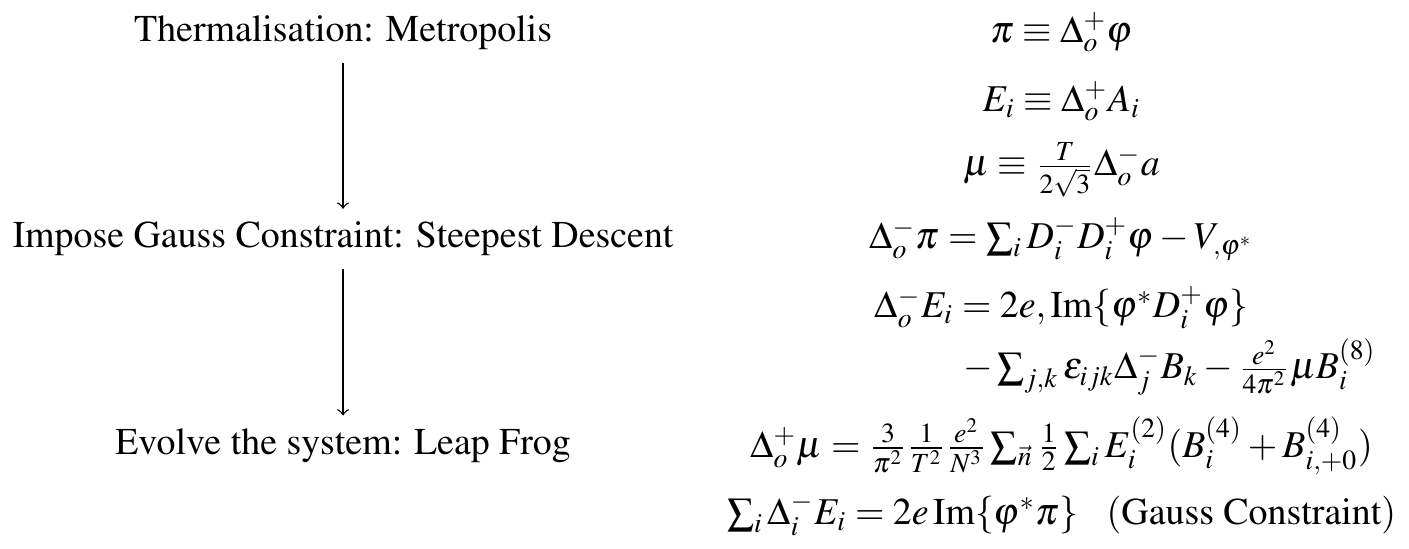};
  \caption{Lattice problem and equations of motion. $D_\mu^\pm f =\pm\frac{1}{\rm dx}(e^{\mp i e {\rm dx^\mu}A_\mu(n\pm\frac{1}{2})} f_{\pm\mu}-f)$  are covariant derivatives.}
  \label{fig:eom}
\end{figure}

To discretise the theory, we fix the temporal gauge $A_0=0$. Given a cubic and periodic lattice of $N^3$ points and lattice spacing $\rm dx$, we provide a discrete complex scalar field $\phi(n)$, a  three components  gauge field $A_i(n+\frac{1}{2}\hat{i})$ together with their conjugate momenta $\pi(n)$ and $E_{i}(n+\frac{1}{2}\hat{i})$. Note that $n$ denotes
a lattice point and $\hat{i}$ a unit vector in the direction $i$; the $\frac{1}{2}$ shift in the gauge fields is reminiscent of their nature as gauge connections. Now the game is to devise a correct discretisation of the topological charge and CS number such that they have the correct lattice continuum limit and obey the correct lattice relations.
Using the notation $f_{a,\mu}=f_a(n+\hat{\mu})$ for $f$ a field with some indices $a$ and denoting $\Delta^{\pm}_\mu f=\pm\frac{1}{\rm dx}(f_{\pm\mu}-f)$ our finite difference operators, we may compute a discrete version of the magnetic field by $B_i(n+\frac{1}{2}\hat{j}+\frac{1}{2}\hat{k})=\sum_{j,k}\epsilon_{ijk}\Delta_j^+A_k$ and then construct the following  operators

\begin{align}
  A_i^{(2)} &\equiv {1\over2}(A_i+A_{i,-i})\ \ & E_i^{(2)} &\equiv {1\over2}(E_i+E_{i,-i})\\
  E_i^{(4)} &\equiv {1\over4}(E_i+E_{i,-i}+E_{i,-0}+E_{i,-i-0})\ &  B_i^{(4)} &\equiv {1\over4}(B_i+B_{i,-j}+B_{i,-k}+B_{i,-j-k})\\
  B_i^{(8)} &\equiv {1\over2}\left(B_i^{(4)}+B_{i,+i}^{(4)}\right)
\end{align}
With these objects we may then define, as promised, a discrete topological charge density $\mathcal{K}^L$ together with a CS density $K^L_\mu$
\begin{align}
  \mathcal{K}^L &\equiv {e^2\over4\pi^2}\sum_i {1\over2}E_i^{(2)}(B_i^{(4)}+B_{i,+0}^{(4)})\\ 
  K_0^L &\equiv -{e^2\over8\pi^2}\sum_i A_{i}^{(2)}B_{i}^{(4)}\\
  K_i^L &\equiv -{e^2\over16\pi^2}\sum_{j,k}\epsilon_{ijk}\left(E_{j}^{(2)}A_{k,-i}^{(2)}+E_{j,-i}^{(2)}A_{k}^{(2)}\right)
\end{align}
Crucially, they satisfy the relation
\begin{equation}
  \mathcal{K}^L=\sum_\mu\Delta^+_\mu K_L^\mu
\end{equation}
With all of this at hand, we can then write down discrete equations of motions, which are shown in figure \ref{fig:eom} and put them together in a real time simulation, which consists in evolving numerically
the equations of motions of some properly conditioned initial field configurations, as summarised in figure \ref{fig:eom}.

\section{Selected Results}
\label{sec:results}
In this section, we discuss some results obtained with these simulations. We focus on the CS diffusion rate and present some initial results on the chemical potential dynamics. For a detailed analysis on the CS rate together with a discussion of the electrical susceptibility
see \cite{Figueroa2018}. A deeper analysis of the chemical potential dynamics will be reported elsewhere \cite{Figueroa2018Boh}.

\subsection{CS Diffusion Rate}
\begin{figure}
  \centering

  \includegraphics{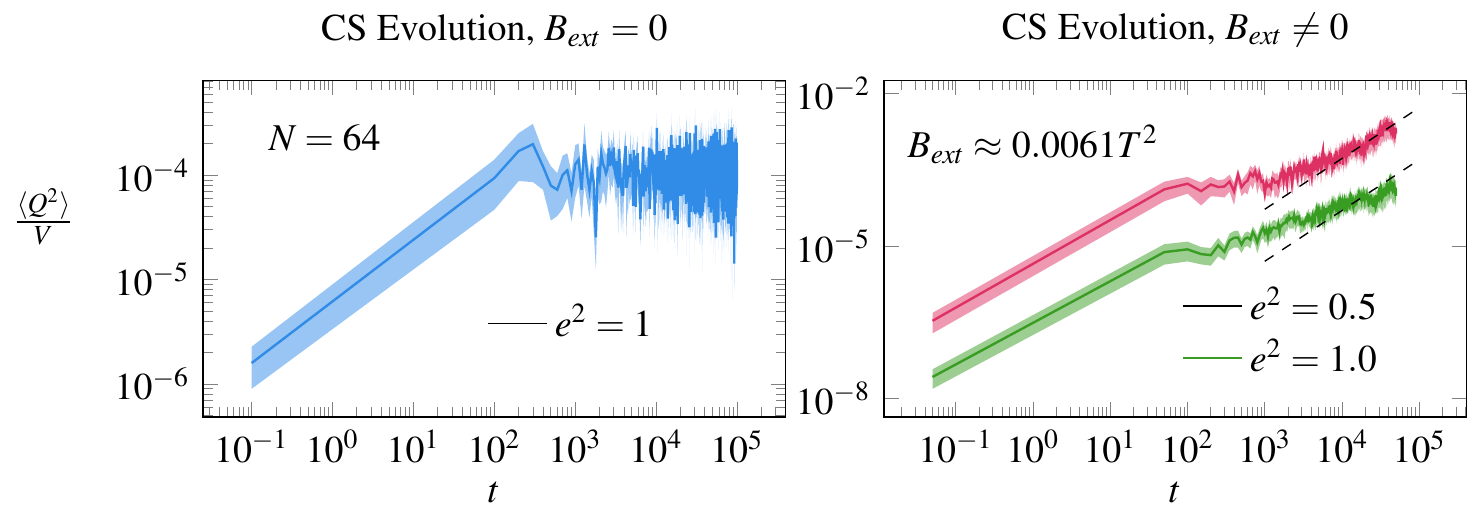};

  \caption{\textbf{Left panel:} In the absence of an external magnetic field, the CS number variance saturates to a constant at late time. \textbf{Right panel:} Magnetically induced diffusion of $\langle Q^2\rangle$.
  Equation \eqref{eq:dif_rate} allows to fit the diffusion coefficients. In both cases, the error bars represent statistical errors.}
  \label{fig:CS}
\end{figure}

Let us start with the diffusion of the CS number. In the absence of external magnetic field, our Abelian theory has a non-degenerate vacuum; creation of CS number costs energy. In other words, for asymptotic times, we expect the CS number to oscillate around zero with a constant variance. This is what we observe in the left hand side of figure \ref{fig:CS}, where we display the variance of the CS number per unit volume. As expected, it asymptotes to a constant value.

When a constant external magnetic field is applied, the situation becomes more interesting \cite{Figueroa2018a}. The external magnetic field introduces a vacuum degeneracy with respect to the CS number and creates a situation very similar to the non-abelian case, where the vacuum degeneracy is intrinsic. This induced degeneracy can be understood
as follow. As $n_{cs}\propto \vec{A}\cdot \vec{B}$, while not requiring any energy, creating a vector potential $\vec{A}$ parallel to $\vec{B}$ changes the CS number; $\vec{B}$ acts as a reservoir of vacua. Thus, $n_{cs}$ will be described as a random walk between vacua and its variance is expected to grow linearly with time. More
precisely, defining $Q(t)\equiv \int dx^3  \left ( n_{cs}(x,t)-n_{cs}(x,0)\right )$
\begin{equation}
  {\langle Q(t)^2\rangle\over V}=\Gamma t
  \label{eq:dif_rate}
\end{equation}
for large $t$ and where $\Gamma$ is the CS diffusion rate. Generically, this rate is expected to depend on the charge of the particles and the strength of the magnetic field. As reported in \cite{Figueroa2018}, one can try to model this dependence in the framework of magnetohydrodynamics (MHD). The dependence is predicted to be of the sort $\Gamma = c_{MHD} e^6 B^2$
with some proportionality factor $c_{MHD}$. In the right hand side panel of figure \ref{fig:CS}, we show the evolution of  ${\langle Q(t)^2\rangle\over V}$. The dashed lines are linear fits of \cite{Figueroa2018} which proves that the late time evolution is indeed linear. This also shows that there is a dependence on the charge. The  full analysis was reported in \cite{Figueroa2018}, where the behaviour
$  \Gamma \propto e^6 B^2$
was confirmed at the $10\%$ level. On the other hand, the measured proportionality factor $c_{measured}$ differs by a factor of order $60$ from the MHD prediction

\begin{equation}
  {c_{measured}\over c_{MHD}}\approx 60
\end{equation}

This result is a strong motivation to pursue these lattice investigations further and study other observables.

\subsection{Chiral Chemical Potential}
Another natural question to ask is: what is the fate of an initial chiral chemical potential? The following simple argument shows that one should expect some instabilities.  In the theory
without fermions, as explained in section \ref{sec:setup}, one adds an effective term $\mu_5 n_{cs}$. At non-zero temperature, the favored state is the one of minimal free energy; there is a competition between the kinetic term and the new effective term. This is best understood in momentum space, where the former is
quadratic in $k$ whereas the latter is linear in $k$. More precisely
\begin{align}
  \frac{1}{2}(E^2+B^2)\sim \frac{1}{2} k^2 A^2, \ \ \ \ \frac{\alpha}{2\pi} \mu_5\vec{A}\cdot \vec{B} \sim\frac{\alpha}{2\pi} \mu_5k A^2
\end{align}
In particular, we see there is an instability whenever
\begin{equation}
  \frac{1}{2} k^2 A^2-\frac{\alpha}{2\pi} \mu_5k A^2<0 \iff k<k_{crit}=\frac{\alpha}{\pi}\mu_5
  \label{eq:kcrit}
\end{equation}
This tells us two things. First, the modes responsible for the instability are in the infrared. The underlying physical picture is the following. As in this theory long-range gauge fields carry less energy than matter, the chiral imbalance is converted in such gauge fields.
Then, we see that this phenomenon will be affected by the lattice infrared cutoff $k_{min}={2\pi \over N}$. Indeed, equation \eqref{eq:kcrit} may be understood as follows: any chemical potential such that
\begin{equation}
  \mu_5< \frac{8 \pi^3} {N e^2}=\mu_c
  \label{eq:mucrit}
\end{equation}
will be stable. In particular, we see that at fixed charge, the smaller the chemical potential we want to study, the larger the lattices need be.

\begin{figure}
  \centering
\includegraphics{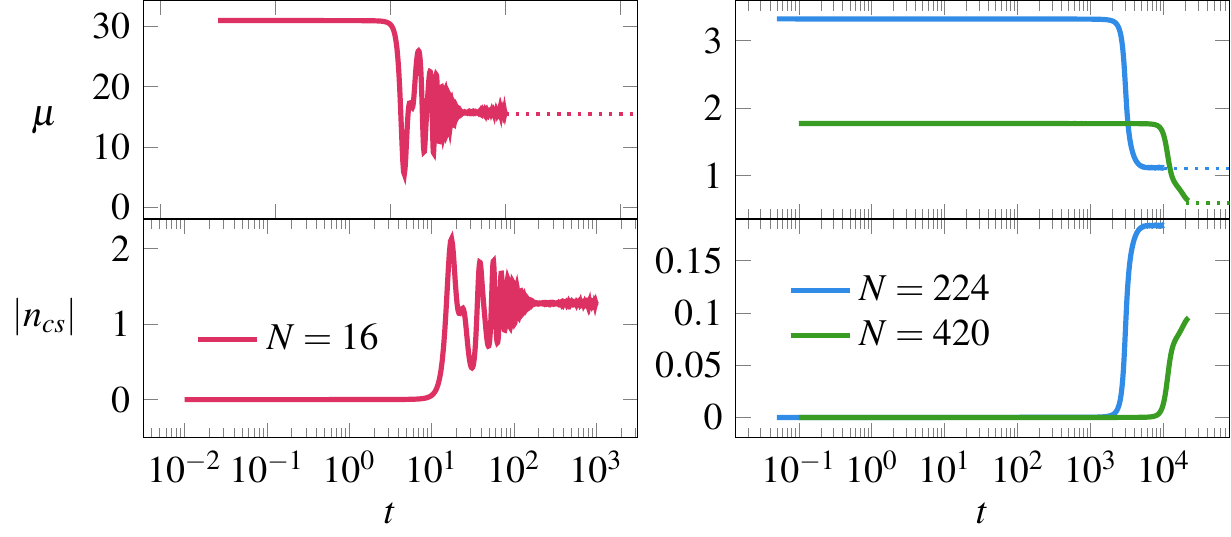};
  \caption{\textbf{Upper panel:} Evolution of $\mu_5$. As expected, it decays into magnetic helicity, which is shown in the \textbf{lower panel}.
  The dotted lines show the predicted value for the lowest chemical potential. Longer plateaus are observed for smaller chemical potentials and damped oscillations for larger ones.}
  \label{fig:mu}
\end{figure}

These predictions are indeed observed on the lattice, as shown in the upper panel of figure \ref{fig:mu}. Initial chemical potentials decay and reach the critical lattice values predicted in equation \eqref{eq:mucrit} (dotted lines). Because of the anomaly equation \eqref{eq:an_rel}, they are transferred into
magnetic helicity, i.e. CS number, see lower panel of figure \ref{fig:mu}. The behaviour strongly depends on the initial value of the chemical potential. For large values (left hand side), the decay seems to proceed at least partly through damped oscillations, even if the full behaviour
cannot be extracted from figure \ref{fig:mu} as the difference between the initial chemical potential and the critical one is of the same order than the amplitude of the oscillations. Going towards smaller chemical potentials (right hand side), we observe a disappearance of the oscillations; the initial charge simply decays to the critical one.
We also encounter a second phenomenon: the smaller the chemical potential, the longer it takes for the decay to be triggered. This makes the investigation of the small chemical potential region even more challenging as it requires large simulations to be run for a long time.
As mentioned earlier, a more complete and quantitative investigation of this dynamic is to be reported elsewhere \cite{Figueroa2018Boh}.

\section{Outlooks}
\label{sec:outlooks}
The results reported here, which shows that while the qualitative feature of the $U(1)$ chiral charge are reasonably well understood, some quantitative estimates are in tension with first principles simulations, giving great motivations to continue this work further. Several directions open up.
On one side, we wish to improve our model, which for now does not take into account hard thermal loops, that turned out to be relevant in the $SU(2)$ case \cite{Arnold1997}. Another direction is to further study the behaviour of the chiral chemical potential. The regime of small
chemical potential is of particular interest. Since gauge fields can have non-zero CS number with arbitrarily small energy, one may hope  to uncover processes similar to non-abelian sphalerons i.e. thermal fluctuations between these quasi-vacua, see section 2.2 in \cite{Figueroa2018a}.

\section*{Acknowledgments}

This work is supported by the Swiss National Science Foundation and realised thanks to the EPFL HPC facility (SCITAS).




\begin{thebibliography}{10}

\bibitem{9780521670548}
S.~Weinberg.
\newblock {\em The Quantum Theory of Fields, Volume 2: Modern Applications}.
\newblock Cambridge University Press, 2005.

\bibitem{Kuzmin1985}
V.~A. Kuzmin, V.~A. Rubakov, and M.~E. Shaposhnikov.
\newblock {On the Anomalous Electroweak Baryon Number Nonconservation in the
  Early Universe}.
\newblock {\em Phys. Lett.}, 155B:36, 1985.

\bibitem{Kharzeev2016}
D.~E. Kharzeev, J.~Liao, S.~A. Voloshin, and G.~Wang.
\newblock {Chiral magnetic and vortical effects in high-energy nuclear
  collisions-A status report}.
\newblock {\em Prog. Part. Nucl. Phys.}, 88:1--28, 2016.

\bibitem{DOnofrio2012a}
M.~D'Onofrio, K.~Rummukainen, and A.~Tranberg.
\newblock {The Sphaleron Rate through the Electroweak Cross-over}.
\newblock {\em JHEP}, 08:123, 2012.

\bibitem{Figueroa2018}
D.~G. Figueroa and M.~E. Shaposhnikov.
\newblock {Anomalous non-conservation of fermion/chiral number in Abelian gauge
  theories at finite temperature}.
\newblock {\em JHEP}, 04:026, 2018.

\bibitem{Figueroa2018a}
D.~G. Figueroa and M.~E. Shaposhnikov.
\newblock {Lattice implementation of Abelian gauge theories with Chern-Simons
  number and an axion field}.
\newblock {\em Nucl. Phys.}, B926:544--569, 2018.

\bibitem{Figueroa2018Boh}
D.~G. Figueroa, A.~Florio, and M.~E. Shaposhnikov.
\newblock To appear.

\bibitem{Rubakov1986}
V.~A. Rubakov.
\newblock {On the Electroweak Theory at High Fermion Density}.
\newblock {\em Prog. Theor. Phys.}, 75:366, 1986.

\bibitem{Moore1996a}
Guy~D. Moore.
\newblock {Motion of Chern-Simons number at high temperatures under a chemical
  potential}.
\newblock {\em Nucl. Phys.}, B480:657--688, 1996.

\bibitem{Arnold1997}
P.~B. Arnold, D.~Son, and L.~G. Yaffe.
\newblock {The Hot baryon violation rate is $O (\alpha_w^5 T^4)$}.
\newblock {\em Phys. Rev.}, D55:6264--6273, 1997.

\end{thebibliography}
\end{document}